\documentclass[aps,prd,superscriptaddress,showpacs,preprint,amsmath,amssymb]{revtex4}
\usepackage{graphicx, bm}
\usepackage[usenames]{color}

\begin{document}

%\tightenlines
\draft
\title{FCC-he sensitivity estimates on the anomalous electromagnetic dipole moments of the top-quark}

\author{ M. A. Hern\'andez-Ru\'{\i}z\footnote{mahernan@uaz.edu.mx}}
\affiliation{\small Unidad Acad\'emica de Ciencias Qu\'{\i}micas, Universidad Aut\'onoma de Zacatecas\\
         Apartado Postal C-585, 98060 Zacatecas, M\'exico.\\}

\author{ A. Guti\'errez-Rodr\'{\i}guez\footnote{alexgu@fisica.uaz.edu.mx}}
\affiliation{\small Facultad de F\'{\i}sica, Universidad Aut\'onoma de Zacatecas\\
         Apartado Postal C-580, 98060 Zacatecas, M\'exico.\\}

\author{M. K\"{o}ksal\footnote{mkoksal@cumhuriyet.edu.tr}}
\affiliation{\small Deparment of Optical Engineering, Sivas Cumhuriyet University, 58140, Sivas, Turkey.\\}

\author{A. A. Billur\footnote{abillur@cumhuriyet.edu.tr}}
\affiliation{\small Deparment of Physics, Sivas Cumhuriyet University, 58140, Sivas, Turkey.\\}

\date{\today}
%\maketitle

\begin{abstract}
% insert abstract here

In this paper, we study the production of a top-quark in association with a bottom-quark and a electron-neutrino
at the Future Circular Collider Hadron Electron (FCC-he) to probe the sensitivity on its magnetic moment $(\hat a_V)$
and its electromagnetic dipole moment $(\hat a_A)$ through the process $e^-p \to e^-\gamma p \to \bar t \nu_e b p$.
Assuming a large amount of collisions, as well as of data with cleaner environments, the FCC-he is an excellent option
to study new physics, such as the $\hat a_V$ and $\hat a_A$. For our sensitivity study on $\hat a_V$ and $\hat a_A$,
we consider center-of-mass energies $\sqrt{s}= 7.07, 10\hspace{0.8mm}TeV$ and luminosities
${\cal L}=50, 100, 300, 500, 1000\hspace{0.8mm}fb^{-1}$. In addition, we apply systematic uncertainties
$\delta_{sys}=0\%, 3\%, 5\%$ and we consider unpolarized and polarized electron beam. Our results show that the
FCC-he is a very good perspective to probe the $\hat a_V$ and $\hat a_A$ at high-energy and high-luminosity frontier.

\end{abstract}


\vspace{5mm}

\maketitle

\section{Introduction}

Electron-proton ($e^-p$) colliders have been and they continue to be considered as ideal machine to probe physics
beyond the Standard Model (BSM). These $e^-p$ colliders, such as the Future Circular Collider Hadron Electron (FCC-he) \cite{FCChe,Fernandez,Fernandez1,Fernandez2,Huan,Acar}
develops options for potential high-energy frontier circular colliders at the post Large Hadron Collider (LHC) era.
Such colliders will open up new perspective in the field of fundamental physics, especially for the particle physics.
Many potential features in favor of this type of electron-proton colliders are the following: 1) Provides a cleaner
environment compared to the $pp$ colliders and higher center-of-mass energies that to the $e^+e^-$ ones. The center-of-mass
energies are much higher than that of the future International Linear Collider (ILC) and the Compact Linear Collider (CLIC).
2) Connection $e^-p$ and $pp$ physics. 3) Provides a cleaner environment with suppressed backgrounds from strong interactions.
4) $e^-p$ when added to $pp$ turns the $pp$ colliders into high precision Higgs and top-quark facilities. Removes the
Parton Distribution Function (PDF) and coupling constant uncertainties in $pp$, $gg$ fusion processes. 5) This collider
could also provides additional and sometimes unique ways for studying the Higgs boson and top-quark physics, as well as the
exploration of Electroweak Symmetry Breaking (EWSB) phenomena, with unmatchable precision and sensitivity. 6) Statistics
enhanced by several orders of magnitude for BMS phenomena brought to light by the LHC. 7) Benefit from both
direct (large $Q^2$) and indirect (precision) probes. 8) Provides solid answers to open questions of the Standard Model (SM)
like: hierarchy problem, prevalence of matter over antimatter, the neutrino masses, dark matter and dark energy. In summary,
$e^-p$ + $pp$ deliver high precision of Higgs boson, top-quark and QCD and electroweak physics complementary to $e^+e^-$.
Furthermore, $e^-p$ is an stimulating, realistic option for a next energy frontier collider for particle physics.
For a comprehensive study on the physics and detector design concepts we refer the readers to Refs.
\cite{FCChe,Fernandez,Fernandez1,Fernandez2,Huan,Acar}.

Next the detection of the top-quark, there has been an enormous motivation to investigate the properties and the potential
of top-quark in great detail both in production and in decay.  Increasingly sophisticated experimental results of the
current colliders are complemented by very precise theoretical predictions in the framework of the SM of
particle physics and beyond. Specifically, the anomalous coupling $t\bar t\gamma$, which is the subject of this paper,
have been studied in detail in hadron colliders and at a future high-luminosity high-energetic linear electron-positron
colliders, for a reviews exhaustive see Table I of Ref. \cite{Billur0} and references therein \cite{Juste,Baur,Bouzas,Bouzas1,Sh,Aguilar,murat,Billur,Sukanta,Seyed,Rindani}.

The SM prediction for the Magnetic Moment (MM) and the Electric Dipole Moment (EDM) of the top-quark, that is $a_t$ \cite{Benreuther}
and $d_t$ \cite{Hoogeveen,Pospelov,Soni} reads:

\begin{equation}
a_t= 0.02,  \hspace{1cm}
d_t < 10^{-30} (e cm).
\end{equation}

\noindent The $a_t$ can be tested in the current and future colliders such as the LHC, CLIC, the Large Hadron-Electron Collider (LHeC)
and the FCC-he. In the case of the $d_t$, its value is strongly suppressed as shown in Eq. (1),
and is much too hard to be observed. However, it is very attractive for probing new physics BSM. Furthermore, it is considered as
a source of CP violation.

With everything already mentioned, the FCC-he potential of $e^-p$ collisions at $\sqrt{s}= 7.07\hspace{0.8mm}TeV$ and $10\hspace{0.8mm}TeV$
and high luminosities ${\cal L}=50-1000\hspace{0.8mm}fb^{-1}$, offer one of the best opportunities to test and improve our understanding
of the top-quark physics. In special their MM and EDM, and as we already mentioned with the latter
considered as a source of CP violation. CP violation can explain why there is more matter than antimatter in the universe,
which is a topic of great relevance between the scientific community of particles and fields.

The MM and EDM of the top-quark, that is $\hat a_V$ and $\hat a_A$ can be probed in high-energy electron-proton collisions
through the neutral current top-quark production, there are mainly two  modes, i) Deep Inelastic Scattering (DIS)
and ii) Photoproduction. Single top-quark and top-quark pair production is possible by both mechanisms.

In this paper we study in a model-independent way the dipole moments of the top-quark through the process of single
top-quark production $e^-p \to e^-\gamma p \to \bar t \nu_e b p$. Fig. 1 shows the schematic diagram for the process
$e^-p \to e^-\gamma p \to \bar t \nu_e b p$, while, the Feynman diagrams contributing to the reaction
$e^-\gamma \to \bar t \nu_e b $ they are shown in Fig. 2.

The rest of the paper is organized as follows. In Section II, we introduce the top-quark effective electromagnetic interactions.
In Section III, we sensitivity estimates on top-quark anomalous electromagnetic couplings through $e^-p \to e^-\gamma p \to \bar t \nu_e b p$
collisions. Finally, we present our conclusions in Section IV.

\section{Single top-quark production via the process $e^-p \to e^-\gamma p \to \bar t \nu_e b p$}

\subsection{Effective interaction of $t\bar t \gamma$}

Due to the absence so far of any signal of new heavy particles decaying into top-quark, an attractive approach for describing
possible new physics effects in a model independent way is based on effective Lagrangian. In this approach, all the heavy
degrees of freedom are integrated out leading to obtain the effective interactions between the SM particles. This is justified
because the related observables have not shown any significant deviation from the SM predictions so far. From the effective
Lagrangian approach, potential deviations of its value from the SM for the anomalous $tt\gamma$ coupling are described of the
effective Lagrangian given by

\begin{equation}
{\cal L}_{eff}={\cal L}_{SM} + \sum_n \frac{c_n}{\Lambda^2}{\cal O}^{(6)}_n + h.c..
\end{equation}

\noindent In Eq. (2), ${\cal L}_{eff}$ is the effective Lagrangian gauge-invariant which contains a series of dimension-six operators
built with the SM fields, ${\cal L}_{SM}$ is the renormalizable SM Lagrangian, $\Lambda$ is the scale at which new physics
expected to be observed, $c_n$ are dimensionless coefficients and ${\cal O}^{(6)}_n$ represents the dimension-six gauge-invariant
operator.

We write the most general effective vertex of $t\bar t\gamma$ \cite{Sh,Kamenik,Baur,Aguilar,Aguilar1} as:

\begin{equation}
{\cal L}_{t\bar t\gamma}=-g_eQ_t\bar t \Gamma^\mu_{ t\bar t  \gamma} t A_\mu,
\end{equation}

\noindent this equation includes the SM coupling and contributions from dimension-six effective operators. In addition,
$g_e$ is the electromagnetic coupling constant, $Q_t$ is the top-quark electric charge and the Lorentz-invariant vertex
function $\Gamma^\mu_{t\bar t \gamma}$ is given by

\begin{equation}
\Gamma^\mu_{t\bar t\gamma}= \gamma^\mu + \frac{i}{2m_t}(\hat a_V + i\hat a_A\gamma_5)\sigma^{\mu\nu}q_\nu.
\end{equation}

\noindent Here $m_t$ is the mass of the top-quark, $q$ is the momentum transfer to the photon and the couplings
$\hat a_V$ and $\hat a_A$ are real and related to the anomalous magnetic moment $(a_t)$ and the electric dipole moment
$(d_t)$ of the top-quark, respectively. The relations between $\hat a_V (\hat a_A) $ and $a_t (d_t)$ are given by

\begin{eqnarray}
\hat a_V&=&Q_t a_t,  \\
\hat a_A&=&\frac{2m_t}{e}d_t.
\end{eqnarray}

The operators contribute to top-quark eletromagnetic anomalous couplings \cite{Buhmuller,Aguilar2,Antonio} are

\begin{eqnarray}
{\cal O}_{uW}^{33}&=&\bar q_{L3}\sigma^{\mu\nu}\tau^a t_{R}{\tilde \phi} W_{\mu\nu}^{a}+{\mbox{h.c}},\\
%\end{eqnarray}
%\begin{eqnarray}
{\cal O}_{uB\phi}^{33}&=&\bar q_{L3}\sigma^{\mu\nu}t_{R}{\tilde \phi} B_{\mu\nu}+{\mbox{h.c}},
\end{eqnarray}

\noindent where $\bar q_{L3}$ is the quark field, $\sigma^{\mu\nu}$ are the Pauli matrices and ${\tilde \phi}$ is the Higgs doublet,
while $W_{\mu\nu}^{a}$ and $B_{\mu\nu}$ are the $U(1)_Y$ and $SU(2)_L$ gauge field strength tensors, respectively.

From the parametrization given by Eq. (3), and from the operators of dimension-six given in Eqs. (7) and (8) we obtain
the corresponding CP even ${\hat a_V}$ and CP odd ${\hat a_A}$ observables:

\begin{eqnarray}
\hat a_V=\frac{2 m_t}{e} \frac{\sqrt{2}\upsilon}{\Lambda^{2}} Re\Bigl[\cos\theta _{W} C_{uB\phi}^{33} + \sin\theta _{W} C_{uW}^{33}\Bigr],\\
%\end{eqnarray}
%\begin{eqnarray}
\hat a_A=\frac{2 m_t}{e} \frac{\sqrt{2}\upsilon}{\Lambda^{2}} Im\Bigl[\cos\theta _{W} C_{uB\phi}^{33} + \sin\theta _{W} C_{uW}^{33}\Bigr].
\end{eqnarray}

\noindent These observables contain $\upsilon=246$ GeV, the breaking scale of the electroweak symmetry and
$\sin\theta _{W} (\cos\theta _{W})$, the sine(cosine) of the weak mixing angle.

\subsection{Cross-section of $e^-p \to e^-\gamma p \to \bar t \nu_e b p$}

The FCC-he will be designed to operate in $e^{-}p$ collision mode, but it can also be operated as a $e^- \gamma$, $\gamma p$ and
$\gamma \gamma$ collider. A promising mechanism to generate energetic photon beams in a FCC-he is Equivalent Photon Approximation (EPA) \cite{Budnev,Baur1,Piotrzkowski} using the Weizsacker-Williams Approximation. In EPA, photons emitted from incoming hadrons (leptons)
which have very low virtuality are scattered at very small angles from the beam pipe. As is know, there is an elastic collision
between the electron and the photon. The emitted quasi-real photons have a low $Q^{2}$ virtuality and are almost real, due to
this reason the proton which emitted photons remain intact and do not dissociate into partons. Intact protons can be detected
by forward detectors almost simultaneously with electron-photon collisions \cite{exp1,exp2,exp3,exp4}. However, the
particles in the final state of $\gamma e$ collisions will go to the central detectors. A process that can be considered as a
background of the process $e^{-}p\rightarrow e^{-} \gamma p \rightarrow \bar{t} b \nu_e p$ at the FCC-he is the process
$e^{-}p\rightarrow e^{-} \gamma p \rightarrow e^-\bar{t} b + X$ as an inelastic process. Such a process may occur
with the $\gamma W^- \to \bar t b$ subprocess and contains only electroweak interactions. In this process, the photon is emitted
by the electron and this electron can be detected by the forward detectors.

First of all, the process $e^{-}p\rightarrow e^{-} \gamma p \rightarrow e^{-}\bar{t} b +X $ is different from the process $e^{-}p\rightarrow e^{-} \gamma p \rightarrow \bar{t} b \nu_e p$,  since the process we examined there is the electron neutrino in the final state. The electron neutrino is not detected directly in the central
detectors. Instead, their presence is inferred from missing energy signal. However, the colliding proton in the process $e^{-}p\rightarrow e^{-} \gamma p \rightarrow e^-\bar{t} b + X$ dissociates into partons. In addition, elastic processes can be distinguished completely inelastic processes due to some experimental signatures.
First, as mentioned above, after the elastic emission of photon, proton is scattered with a small angle and escapes detection
from the central detectors. This gives rise to a missing energy signature called forward large-rapidity gap, in the corresponding
forward region of the central detector. However, productions $\ell \bar{\ell}$, $\gamma\gamma$, $jj$ and $J/\psi$ of elastic
processes with the aid of this technique were successfully examined by CDF and CMS Collaborations \cite{fer1,fer2,fer3,cms,cms1}.
Also, another experimental signature can be implemented by forward particle tagging. These detectors are to tag the forward
protons with some energy fraction loss. CMS and TOTEM Collaborations at the LHC began these measurements using forward detectors
between the CMS interaction point and detectors in the TOTEM area about 210 m away on both sides of interaction point \cite{1}.
However, LHeC Collaboration has a program of forward physics with extra detectors located in a region between a few tens up to
several hundreds of meters from the interaction point \cite{2}. For these reasons, the $\gamma e $ collisions in the correlation
with the central detectors can be determined as a separate signal from the $\gamma p$ collisions.

In this paper our calculations are based on photon-electron fluxes through the subprocess $e^-\gamma \to \bar t \nu_e b $ and the
representative leading order Feynman diagrams are depicted in Fig. 2. Here we put in evidence the contribution of elastic process
with an intact proton in the final state, as well as the inelastic component for the leptonic final state. The spectrum of EPA photons
which are emitted by proton is given by \cite{Belyaev,Budnev}

\begin{eqnarray}
f_{\gamma}(x)=\frac{\alpha}{\pi E_{p}}\{[1-x][\varphi(\frac{Q_{max}^{2}}{Q_{0}^{2}})-\varphi(\frac{Q_{min}^{2}}{Q_{0}^{2}})],
\end{eqnarray}

\noindent where $x=E_{\gamma}/E_{p}$ and $Q^2_{max}$ is the maximum virtuality of the photon. In our calculations, we use
$Q^2_{max}=2\hspace{0.8mm}GeV^2$. The minimum value of the $Q^2_{min}$ is given by

\begin{eqnarray}
Q_{min}^{2}=\frac{m_{p}^{2}x^{2}}{1-x}.
\end{eqnarray}

From Eq. (11), the function $\varphi$ is the following

\begin{eqnarray}
\varphi(\theta)=&&(1+ay)\left[-\textit{In}(1+\frac{1}{\theta})+\sum_{k=1}^{3}\frac{1}{k(1+\theta)^{k}}\right]
+\frac{y(1-b)}{4\theta(1+\theta)^{3}} \nonumber \\
&& +c(1+\frac{y}{4})\left[\textit{In}\left(\frac{1-b+\theta}{1+\theta}\right)+\sum_{k=1}^{3}\frac{b^{k}}{k(1+\theta)^{k}}\right],
\end{eqnarray}

\noindent where explicitly $y$, $a$, $b$ and $c$ are as follows

\begin{eqnarray}
y=\frac{x_{2}^{2}}{(1-x_{2})},
\end{eqnarray}

\begin{eqnarray}
a=\frac{1+\mu_{p}^{2}}{4}+\frac{4m_{p}^{2}}{Q_{0}^{2}}\approx 7.16,
\end{eqnarray}

\begin{eqnarray}
b=1-\frac{4m_{p}^{2}}{Q_{0}^{2}}\approx -3.96,
\end{eqnarray}

\begin{eqnarray}
c=\frac{\mu_{p}^{2}-1}{b^{4}}\approx 0.028.
\end{eqnarray}

Hence, the total cross-section of the scattering $e^-p \to e^-\gamma p \to \bar t \nu_e b p$ can be expressed as

\begin{eqnarray}
\sigma(e^-p \to e^-\gamma p \to \bar t \nu_e b p)=\int f_{\gamma}(x)d\hat{\sigma}( e^-\gamma \to \bar t \nu_e b) dx,
\end{eqnarray}

\noindent where $\sigma( e^-\gamma \to \bar t \nu_e b)$ is the cross-section of the scattering $e^-\gamma \to \bar t \nu_e b$
and $f_{\gamma}(x)$ is the spectrum of equivalent photons which is given in Eq. (11).

It is worth mentioning that, there are different ways to optimize the signal sensitivity $e^-p \to e^-\gamma p \to \bar t \nu_e b p$
and reduce the background. This is possible if we apply cut-based optimization, in addition to considering polarized electron beam.

We base our results on the following kinematic acceptance cuts in order to optimize the significance of the signal over all the backgrounds:

\begin{eqnarray}
\begin{array}{c}
\noindent \mbox{Cut-1:} \hspace{2mm} p^b_T > 20\hspace{0.8mm}GeV,
\hspace{1cm} \mbox{Cut-2:} \hspace{2mm} |\eta^b| < 2.5,  \hspace{1cm}
\mbox{Cut-3:} \hspace{2mm} p^{\nu_e}_T > 15\hspace{0.8mm}GeV.
\end{array}
\end{eqnarray}

In Eq. (19), $p^b_T$ is the transverse momentum of the final state bottom-quark, $\eta^b$ is the pseudorapidity and $p^{\nu_e}_T$
is the transverse momentum of the electron-neutrino. The outgoing particles are required to satisfy these isolation cuts.

An essential feature in the design of current and future colliders of high-energy physics, is the implementation of polarized
particles beams. Most accelerators have been modified or are being designed with the possibility of using polarized particles
sources, such as the FCC-he. The possibility of using polarized electron beams can constitute a strong advantage in searching
for new physics \cite{Moortgat}. Furthermore, the electron beam polarization may lead to a reduction of the measurement
uncertainties, either by increasing the signal cross-section, therefore reducing the statistical uncertainty, or by suppressing
important backgrounds. In summary, one another option at the FCC-he is to polarize the incoming beam, which could maximize the
physics potential, both in the performance of precision tests and in revealing the properties of the new physics BSM.

The general formula for the total cross-section for an arbitrary degree of longitudinal $e^-$ and $e^+$ beams polarization is given
by \cite{Moortgat}

\begin{eqnarray}
\sigma(P_{e^-},P_{e^+})=&&\frac{1}{4}[(1+P_{e^-})(1+P_{e^+})\sigma_{++}+(1-P_{e^-})(1-P_{e^+})\sigma_{--}\nonumber\\
&&+(1+P_{e^-})(1-P_{e^+})\sigma_{+-}+(1-P_{e^-})(1+P_{e^+})\sigma_{-+}],
\end{eqnarray}

\noindent where $P_{e^-} (P_{e^+})$ is the polarization degree of the electron (positron) beam, while $\sigma_{-+}$
stands for the cross-section for completely left-handed polarized $e^-$ beam $P_{e^-}=-1$ and completely right-handed
polarized $e^+$ beam $P_{e^+}=1$, and other cross-sections $\sigma_{--}$, $\sigma_{++}$ and $\sigma_{+-}$ are defined
analogously.

The main anomalous electromagnetic couplings affecting top-quark physics that are of interest for our study are
$\hat a_V$ and $\hat a_A$. We have calculated the dependencies of the $e^-p \to e^-\gamma p \to \bar t \nu_e b p$
production cross-sections for the FCC-he at $7.07\hspace{0.8mm}TeV$ and $10\hspace{0.8mm}TeV$ on $\hat a_V$
and $\hat a_A$ using CalcHEP \cite{Belyaev}. Furthermore, for our study we consider unpolarized and polarized electron
beam, as well as the basic acceptance cuts given in Eq. (19). Assuming only one anomalous coupling to be non-zero at at
time, we obtain the following results for the total cross-section in terms of the dipole moments of the top-quark:\\

$i)$ Total cross-section for $\sqrt{s}=7.07\hspace{0.8mm} TeV$ and $P_{e^-}=0\%$.

\begin{eqnarray}
\sigma(\hat a_V)&=&\Bigl[(0.00199)\hat a^2_V +(0.0000350)\hat a_V  +0.000522 \Bigr] (pb),   \\
\sigma(\hat a_A)&=&\Bigl[(0.00199)\hat a^2_A + 0.000522 \Bigr] (pb).
\end{eqnarray}

$ii)$ Total cross-section for $\sqrt{s}=10\hspace{0.8mm} TeV$ and $P_{e^-}=0\%$.

\begin{eqnarray}
\sigma(\hat a_V)&=&\Bigl[(0.00499)\hat a^2_V +(0.0000217)\hat a_V  +0.000777 \Bigr] (pb),   \\
\sigma(\hat a_A)&=&\Bigl[(0.00499)\hat a^2_A + 0.000777 \Bigr] (pb).
\end{eqnarray}

$iii)$ Total cross-section for $\sqrt{s}=7.07\hspace{0.8mm} TeV$ and $P_{e^-}=-80\%$.

\begin{eqnarray}
\sigma(\hat a_V)&=&\Bigl[(0.00358)\hat a^2_V +(0.0000614)\hat a_V  +0.00094 \Bigr] (pb),   \\
\sigma(\hat a_A)&=&\Bigl[(0.00358)\hat a^2_A + 0.00094 \Bigr] (pb).
\end{eqnarray}

$iv)$ Total cross-section for $\sqrt{s}=10\hspace{0.8mm} TeV$ and $P_{e^-}=-80\%$.

\begin{eqnarray}
\sigma(\hat a_V)&=&\Bigl[(0.00898)\hat a^2_V +(0.000037)\hat a_V  +0.0014 \Bigr] (pb),   \\
\sigma(\hat a_A)&=&\Bigl[(0.00898)\hat a^2_A + 0.0014 \Bigr] (pb).
\end{eqnarray}

We see that the sensitivities on the total cross-section and on the coefficients of $\hat a_V$ and $\hat a_A$ increase
with the centre-of-mass energy, as well as with the polarized electron beam, confirming the expected competitive advantage
of the high-energies attainable with the FCC-he.

We first present the total cross-section of the signal $e^-p \to e^-\gamma p \to \bar t \nu_e b p$ as a function of the $\hat a_V$
and $\hat a_A$ for the center-of-mass energies of the FCC-he, that is $\sqrt{s}=7.07\hspace{0.8mm}TeV$ and $\sqrt{s}=10\hspace{0.8mm}TeV$,
as shown through Figs. 3-6. These results are obtained after applying the kinematic cuts given in Eq. (19) and with unpolarized electron
beam $P_{e^-}=0\%$. The results show a clear dependence of the total cross-section of the $e^-p \to e^-\gamma p \to \bar t \nu_e b p$
scattering with respect to $\hat a_V$ and $\hat a_A$, as well as with the center-of-mass energies of the FCC-he.

In the case of the cross-section of the photo-production process $e^-p \to e^-\gamma p \to \bar t \nu_e b p$ after application of cuts
given by Eq. (19) and with polarized electron beam $P_{e^-}=-80\%$, the total cross-section is about 1.8 times larger than that of the
photo-production process $e^-p \to e^-\gamma p \to \bar t \nu_e b p$ with unpolarized electron beam $P_{e^-}=0\%$, as shown in Figs. 9-12.

Before continuing with our study, it is worth making a discussion about our results obtained in Fgs. 3-6 and 9-12.
While the theory predictions for $\hat a_V$ and $\hat a_A$ in Eqs. (5) and (6) as well as the total cross-section that contains the anomalous
coupling $t\bar t\gamma$ have been made in different contexts (see Table I of Ref. \cite{Billur0}), the $\hat a_V$ and $\hat a_A$ have not been
measured experimentally yet. Therefore, one only has the option of comparing the measured $\hat a_V$ and $\hat a_A$ and the total cross-section
with the theoretical calculation of Refs. \cite{Bouzas1,Sh}. The authors of Ref. \cite{Bouzas1} specifically measure $\sigma(\gamma e^- \to t\bar t)$
with $10\%$ $(18\%)$ error obtaining the following
results for the MM and the EDM of the top-quark at the LHeC: $|\kappa|< 0.05 (0.09)$ and $|\tilde\kappa|< 0.20 (0.28)$. While
in our case, with the process $e^-p \to e^-\gamma p \to \bar t \nu_e b p$, we obtain: $\hat a_V=(-0.1480, 0.1438)$ and $\hat a_A=|0.1462|$
with $\sqrt{s}=10\hspace{0.8mm}TeV$, ${\cal L}=1000\hspace{0.8mm}fb^{-1}$, $\delta_{sys}=5\%$, $P_{e^-}=0\%$ and $95\%\hspace{0.8mm}C.L.$.
With polarized electron beam $P_{e^-}=-80\%$, we obtain: $\hat a_V=(-0.1394, 0.1353)$ and $\hat a_A=|0.1374|$. Although the conditions for
the study of both processes $\gamma e^- \to t\bar t$ and $e^-p \to e^-\gamma p \to \bar t \nu_e b p$ are different, our result are competitive
with respect to the results reported in Ref. \cite{Bouzas1}. In addition, it should be noted that our results are for $95\%\hspace{0.8mm}C.L.$,
while those reported in Ref. \cite{Bouzas1} are for $90\%\hspace{0.8mm}C.L.$. On the other hand, from the comparison of our result using the
process $e^-p \to e^-\gamma p \to \bar t \nu_e b p$ at the FCC-he, with respect to the process $pp\to p\gamma^* \gamma^* p \to pt\bar t p$
at LHC, our results show a significant improvement. Furthermore, it is noteworthy that with our process the total cross-sections is a factor
${\cal O}(10^3)$ between $pp\to p\gamma^* \gamma^* p \to pt\bar t p$ and $e^-p \to e^-\gamma p \to \bar t \nu_e b p$, that is, our results
project 3 orders of magnitude better than those reported in Ref. \cite{Sh}. These projections shows that the sensitivity on the anomalous
couplings $\hat a_V$ and $\hat a_A$ can be improved at the FCC-he by a few orders of magnitude in comparison with the projections of the LHC.

\section{Model-independent sensitivity estimates on the $\hat a_V$ and $\hat a_A$}

To determine the sensitivity of the non-standard couplings, $\hat a_V$ and $\hat a_A$, Eqs. (9) and (10) we use the results
from Section II, for the process $e^-p \to e^-\gamma p \to \bar t \nu_e b p$ at the FCC-he. We consider the center-of-mass
energies $\sqrt{s}=7.07, 10\hspace{0.8mm}TeV$ and luminosities ${\cal L}= 50, 100, 300, 500, 1000\hspace{0.8mm}fb^{-1}$ with
unpolarized and polarized electron beam. Furthermore, we consider the kinematic acceptance cuts given by Eq. (19), take into
account the systematic uncertainties $\delta_{sys}=0\%, 3\%, 5\%$ and we follow three different confidence level (C.L.) $68\%$,
$90\%$ and $95\%$ and to make our study more effective we perform a $\chi^2$ test define as:

\begin{equation}
\chi^2(\hat a_V, \hat a_A)=\biggl(\frac{\sigma_{SM}-\sigma_{BSM}(\sqrt{s}, \hat a_V, \hat a_A)}{\sigma_{SM}\sqrt{(\delta_{st})^2
+(\delta_{sys})^2}}\biggr)^2.
\end{equation}

\noindent Here $\sigma_{SM}$ is the cross-section from the SM, while $\sigma_{BSM}(\sqrt{s}, \hat a_V, \hat a_A)$ is the total
cross-section which contains contributions from the SM, as well as non-standard contributions which come from the anomalous
couplings $\hat a_V$ and $\hat a_A$. $\delta_{st}=\frac{1}{\sqrt{N_{SM}}}$ and $\delta_{sys}$ are the statistical and systematic
uncertainties. In our study we consider $\delta_{sys}=0\%, 3\%, 5\%$. The number of events for the process
$e^-p \to e^-\gamma p \to \bar t \nu_e b p$ is given by $N_{SM}={\cal L}_{int}\times \sigma_{SM} \times BR \times \epsilon_{b-tag}$,
where ${\cal L}_{int}$ is the integrated FCC-he luminosity and  $b$-jet tagging efficiency is $\epsilon_b=0.8$ \cite{atlas}.
The top-quark decay almost $100\%$ to $W$ boson and $b$ quark, specifically $\bar t\to \bar bW^-$, where the $W$ boson decays
into leptons and hadrons.

The $\chi^2(\hat a_V, \hat a_A)$ analysis due systematic uncertainties is studied for three representative values of $\delta_{sys}$
at $0\%$, $3\%$ and $5\%$, respectively. And the sensitivity of $\hat a_V$ and $\hat a_A$ at $95\%$ C.L. is found to be of the order
of $10^{-1}$ with $\sqrt{s}=10\hspace{0.8mm}TeV$, ${\cal L}= 1000\hspace{0.8mm}fb^{-1}$ and we consider both cases, that is, unpolarized
and polarized electron beam, as shown in Table VI (which includes the acceptance cuts, Eq. (19)). The order of the sensitivity on the
anomalous couplings $\hat a_V$ and $\hat a_A$ for other values of $\sqrt{s}$ and ${\cal L}$ varies between $10^{-2}-10^{-1}$ at $68\%$ C.L.
and $90\%$ C.L., as shown in Tables I-V. Our study shows that the anomalous $t\bar t \gamma$ vertex at the FCC-he can be probed to a very
good accuracy and is comparable with others existing limits, see Table I, Ref. \cite{Billur0}.

Figs. 7-8 (unpolarized electron beam) and 13-14 (polarized electron beam) show the $95\%$ C.L. contours for the anomalous top-quark
dipole couplings $\hat a_V$ and $\hat a_A$ with the assumed energies and luminosities of $\sqrt{s}=7.07, 10\hspace{0.8mm}TeV$ and
${\cal L}= 50, 250, 1000\hspace{0.8mm}fb^{-1}$. With the uncertainty of $0\%$, the $95\%$ C.L. sensitivity on the couplings are found
to be $\hat a_V  \hspace{1mm}\in \hspace{1mm}[-0.45, 0.05]$, $\hat a_A \hspace{1mm}\in \hspace{1mm}[-0.25, 0.25]$ with $P_{e^-}=0\%$,
and $\hat a_V  \hspace{1mm}\in \hspace{1mm}[-0.120, 0.120]$, $\hat a_A \hspace{1mm}\in \hspace{1mm}[-0.120, 0.120]$ with with $P_{e^-}=-80\%$.
Of the relations given by Eqs. (5) and (6), the sensitivities on the anomalous dipole moments of the top-quark $\hat a_V$ and $\hat a_A$
are corresponding to the following sensitivities on the magnetic and electric dipole moments of the top-quark:

\begin{equation}
\mbox{$P_{e^-}=0\%$}:
\begin{array}{ll}
-0.675 \leq a_t \leq 0.675, \hspace{3mm}  & \mbox{$95\%$ C.L.}, \\
-1.433 \leq d_t (10^{-17}e cm) \leq 1.433, \hspace{3mm}  & \mbox{$95\%$ C.L.},
\end{array}
%\right.
\end{equation}

\noindent and

\begin{equation}
\mbox{$P_{e^-}=-80\%$}:
\begin{array}{ll}
-0.180 \leq a_t \leq 0.180, \hspace{3mm}  & \mbox{$95\%$ C.L.}, \\
-6.878 \leq d_t (10^{-18}e cm) \leq 6.878, \hspace{3mm}  & \mbox{$95\%$ C.L.}.
\end{array}
%\right.
\end{equation}

For the anomalous magnetic and electric dipole moments, an improvement is reachable in comparison with the constraints obtained from
the $\gamma e^- \to t\bar t$ \cite{Bouzas1} and $pp\to p\gamma^* \gamma^* p \to pt\bar t p$ \cite{Sh} searches mentioned previously.

\section{Conclusions}

In this paper, we have study feasibility of measuring the non-standard couplings $\hat a_V$ and $\hat a_A$ coming from the
effective electromagnetic interaction $t\bar t \gamma$ through the process $e^-p \to e^-\gamma p \to \bar t \nu_e b p$ at
the FCC-he. Specifically, we assume energies from 7.07 and 10 TeV and integrated luminosities of at least 50, 100, 300, 5000
and 1000 $fb^{-1}$. Further our sensitivity study is cut-based, polarized electron beam and sources of systematic uncertainties
such as leptons and $b$-jet identification, as well as in a $\chi^2(\hat a_V, \hat a_A)$ test to extract, enhance and optimize
the expected signal cross-section and the sensitivity on $\hat a_V$ and $\hat a_A$. We find that the total cross-section
$\sigma(e^-p \to e^-\gamma p \to \bar t \nu_e b p)$ has a strong dependence on the anomalous couplings $\hat a_V$ and $\hat a_A$,
as well as with the center-of-mass energies of the FCC-he and therefore strong sensitivity estimated are obtained on
$\sigma(e^-p \to e^-\gamma p \to \bar t \nu_e b p)$ (see Figs. 3-6 and 9-12) and $\hat a_V$ ($\hat a_A$) (see Tables I-VI).
Therefore, our results show that with the process $e^-p \to e^-\gamma p \to \bar t \nu_e b p$ at the FCC-he, the sensitivity
estimated on the MM and the EDM of the top-quark can be significantly strengthened. Specifically, with $1000\hspace{0.8mm}fb^{-1}$
of data, $\sqrt{s}=10\hspace{0.8mm}TeV$, $\delta_{sys}=5\%$ and $P_{e^-}=-80\%$ we obtain: $\hat a_V=(-0.1394, 0.1353)$ and
$\hat a_A=|0.1374|$. Our results are competitive with those results shown in Table I of Ref. \cite{Billur0}. At this time, the
FCC-he is an excellent option for the electron-proton collider. It will be useful for any new physics study. Fortunately,
future of $e^-p$ colliders remain promising as it is a natural option like a hybrid between the hadron pp and linear $e^+e^-$ colliders.

It is worth mentioning that, additional improvements could be achieved on the observables of the top-quark, especially in their
electromagnetic properties to the extent that more sophisticated analysis methods are apply. In addition to the improvement in
the technology of detection of the current and future high-energy physics colliders.

%\tabla 2
\begin{table}
\caption{Sensitivity on the $\hat a_V$ magnetic moment and the $\hat a_A$ electric dipole moment of the top-quark
through the process $e^-p \to e^-\gamma p \to \bar t \nu_e b p$.}
\begin{center}
\begin{tabular}{|cccccc|}
\hline\hline
\multicolumn{6}{|c|}{$\sqrt{s}=7.07$ TeV, \hspace{1cm}   $68\%$ C.L.}\\
 \hline
 & & \multicolumn{2}{|c|}{$P_{e^-}=0\%$} & \multicolumn{2}{|c|}{$P_{e^-} = -80\%$}\\
 \hline
 \cline{1-6}
 \hspace{0.5cm} ${\cal L}\, (fb^{-1})$  & \hspace{0.5cm} $ \delta_{sys}$ \hspace{0.5cm}  &  $\hat a_V$   & \hspace{0.5cm} $|\hat a_A|$  \hspace{0.5cm}  & $\hat a_V$   & \hspace{0.5cm} $|\hat a_A|$ \hspace{0.5cm} \\
\hline
50  &  $0\%$   & [-0.2733, 0.2557]   &    0.2644  & [-0.2370, 0.2199]  & 0.2282  \\
50  &  $3\%$   & [-0.2742, 0.2566]   &    0.2652  & [-0.2383, 0.2212]  & 0.2295  \\
50  &  $5\%$   & [-0.2756, 0.2580]   &    0.2667  & [-0.2406, 0.2234]  & 0.2318  \\
\hline
100 &  $0\%$   & [-0.2313, 0.2137]    &    0.2223  & [-0.2007, 0.1836]  & 0.1919 \\
100 &  $3\%$   & [-0.2327, 0.2151]    &    0.2237  & [-0.2029, 0.1857]  & 0.1941 \\
100 &  $5\%$   & [-0.2351, 0.2175]    &    0.2261  & [-0.2066, 0.1894]  & 0.1978 \\
\hline
300  &  $0\%$   & [-0.1779, 0.1603]    &   0.1689  & [-0.1547, 0.1375]  & 0.1458 \\
300  &  $3\%$   & [-0.1811, 0.1635]    &   0.1720  & [-0.1594, 0.1423]  & 0.1506 \\
300  &  $5\%$   & [-0.1862, 0.1686]    &   0.1772  & [-0.1669, 0.1498]  & 0.1581 \\
\hline
500 &  $0\%$   & [-0.1577, 0.1401]    &    0.1487  & [-0.1372, 0.1201]  & 0.1283 \\
500 &  $3\%$   & [-0.1622, 0.1446]    &    0.1532  & [-0.1440, 0.1268]  & 0.1351 \\
500 &  $5\%$   & [-0.1694, 0.1518]    &    0.1603  & [-0.1540, 0.1368]  & 0.1451 \\
\hline
1000 &  $0\%$   & [-0.1341, 0.1165]    &   0.1250  & [-0.1169, 0.0997]  & 0.1079 \\
1000 &  $3\%$   & [-0.1414, 0.1238]    &   0.1323  & [-0.1275, 0.1103]  & 0.1186 \\
1000 &  $5\%$   & [-0.1519, 0.1343]    &   0.1429  & [-0.1414, 0.1242]  & 0.1325 \\
\hline\hline
\end{tabular}
\end{center}
\end{table}

%\tabla 3
\begin{table}[!ht]
\caption{Sensitivity on the $\hat a_V$ magnetic moment and the $\hat a_A$ electric dipole moment of the top-quark
through the process $e^-p \to e^-\gamma p \to \bar t \nu_e b p$.}
\begin{center}
 \begin{tabular}{|cccccc|}
\hline\hline
\multicolumn{6}{|c|}{ $\sqrt{s}=7.07$ TeV,  \hspace{1cm}  $90\%$ C.L.}\\
 \hline
 & & \multicolumn{2}{|c|}{$P_{e^-}=0\%$} & \multicolumn{2}{|c|}{$P_{e^-} = -80\%$}\\
 \hline
 \cline{1-6} \hspace{0.5cm} ${\cal L}\, (fb^{-1})$  & \hspace{0.5cm} $\delta_{sys}$ \hspace{0.5cm} &  $\hat a_V$  & $\hspace{0.5cm} |\hat a_A|$ \hspace{0.5cm}  & $\hat a_V$ & \hspace{0.5cm} $|\hat a_A|$ \hspace{0.5cm} \\
\hline
50  &  $0\%$   & [-0.3084, 0.2908]    &    0.2995  & [-0.2673, 0.2502] & 0.2585  \\
50  &  $3\%$   & [-0.3093, 0.2917]    &    0.3004  & [-0.2688, 0.2516] & 0.2600  \\
50  &  $5\%$   & [-0.3110, 0.2934]    &    0.3021  & [-0.2713, 0.2542] & 0.2626  \\
\hline
100 &  $0\%$   & [-0.2608, 0.2432]    &    0.2518  & [-0.2262, 0.2090] & 0.2174  \\
100 &  $3\%$   & [-0.2624, 0.2447]    &    0.2534  & [-0.2286, 0.2115] & 0.2199  \\
100 &  $5\%$   & [-0.2651, 0.2475]    &    0.2562  & [-0.2328, 0.2156] & 0.2240  \\
\hline
300 &  $0\%$   & [-0.2003, 0.1827]    &    0.1913  & [-0.1740, 0.1569] & 0.1652  \\
300 &  $3\%$   & [-0.2039, 0.1863]    &    0.1949  & [-0.1794, 0.1623] & 0.1706  \\
300 &  $5\%$   & [-0.2097, 0.1921]    &    0.2008  & [-0.1879, 0.1707] & 0.1791  \\
\hline
500 &  $0\%$   & [-0.1774, 0.1598]    &    0.1684  & [-0.1542, 0.1371] & 0.1454  \\
500 &  $3\%$   & [-0.1825, 0.1649]    &    0.1735  & [-0.1619, 0.1448] & 0.1531  \\
500 &  $5\%$   & [-0.1906, 0.1730]    &    0.1816  & [-0.1732, 0.1561] & 0.1644  \\
\hline
1000 &  $0\%$   & [-0.1507, 0.1331]    &    0.1416 & [-0.1311, 0.1140] & 0.1222  \\
1000 &  $3\%$   & [-0.1589, 0.1413]    &    0.1498 & [-0.1432, 0.1260] & 0.1343  \\
1000 &  $5\%$   & [-0.1708, 0.1532]    &    0.1618 & [-0.1589, 0.1418] & 0.1501  \\
\hline\hline
\end{tabular}
\end{center}
\end{table}

%\tabla 4
\begin{table}[!ht]
\caption{Sensitivity on the $\hat a_V$ magnetic moment and the $\hat a_A$ electric dipole moment of the top-quark
through the process $e^-p \to e^-\gamma p \to \bar t \nu_e b p$.}
\begin{center}
 \begin{tabular}{|cccccc|}
\hline\hline
\multicolumn{6}{|c|}{ $\sqrt{s}=7.07$ TeV,  \hspace{1cm}  $95\%$ C.L.}\\
 \hline
 & & \multicolumn{2}{|c|}{$P_{e^-}=0\%$} & \multicolumn{2}{|c|}{$P_{e^-} = -80\%$}\\
 \hline
 \cline{1-6} \hspace{0.5cm} ${\cal L}\, (fb^{-1})$  & \hspace{0.5cm} $\delta_{sys}$ \hspace{0.5cm} &  $\hat a_V$  & $\hspace{0.5cm} |\hat a_A|$ \hspace{0.5cm}  & $\hat a_V$ & \hspace{0.5cm} $|\hat a_A|$ \hspace{0.5cm} \\
\hline
50  &  $0\%$   & [-0.3790, 0.3614]    &    0.3701  &  [-0.3283, 0.3111] &  0.3195  \\
50  &  $3\%$   & [-0.3802, 0.3626]    &    0.3713  &  [-0.3301, 0.3130] &  0.3213  \\
50  &  $5\%$   & [-0.3822, 0.3646]    &    0.3734  &  [-0.3333, 0.3161] &  0.3245  \\
\hline
100 &  $0\%$   & [-0.3201, 0.3025]    &    0.3112  &  [-0.2775, 0.2603] &  0.2687  \\
100 &  $3\%$   & [-0.3221, 0.3045]    &    0.3132  &  [-0.2805, 0.2633] &  0.2717  \\
100 &  $5\%$   & [-0.3255, 0.3079]    &    0.3166  &  [-0.2856, 0.2685] &  0.2768  \\
\hline
300  &  $0\%$   & [-0.2454, 0.2278]   &    0.2365  &  [-0.2130, 0.1958] &  0.2041  \\
300  &  $3\%$   & [-0.2498, 0.2322]   &    0.2409  &  [-0.2196, 0.2025] &  0.2108  \\
300  &  $5\%$   & [-0.2570, 0.2394]   &    0.2481  &  [-0.2301, 0.2130] &  0.2213  \\
\hline
500 &  $0\%$   & [-0.2171, 0.1995]    &    0.2081  &  [-0.1885, 0.1713] &  0.1797  \\
500 &  $3\%$   & [-0.2234, 0.2058]    &    0.2144  &  [-0.1980, 0.1808] &  0.1892  \\
500 &  $5\%$   & [-0.2334, 0.2158]    &    0.2245  &  [-0.2120, 0.1948] &  0.2032  \\
\hline
1000 &  $0\%$   & [-0.1840, 0.1664]   &    0.1750  &  [-0.1599, 0.1428] &  0.1511  \\
1000 &  $3\%$   & [-0.1942, 0.1766]   &    0.1852  &  [-0.1748, 0.1577] &  0.1660  \\
1000 &  $5\%$   & [-0.2090, 0.1914]   &    0.2000  &  [-0.1943, 0.1771] &  0.1855  \\
\hline\hline
\end{tabular}
\end{center}
\end{table}

% Tabla 5
\begin{table}[!ht]
\caption{Sensitivity on the $\hat a_V$ magnetic moment and the $\hat a_A$ electric dipole moment of the top-quark
through the process $e^-p \to e^-\gamma p \to \bar t \nu_e b p$.}
\begin{center}
 \begin{tabular}{|cccccc|}
\hline\hline
\multicolumn{6}{|c|}{ $\sqrt{s}=10$ TeV,  \hspace{1cm}  $68\%$ C.L.}\\
 \hline
 & & \multicolumn{2}{|c|}{$P_{e^-}=0\%$} & \multicolumn{2}{|c|}{$P_{e^-} = -80\%$}\\
 \hline
 \cline{1-6} \hspace{0.5cm} ${\cal L}\, (fb^{-1})$  & \hspace{0.5cm} $\delta_{sys}$ \hspace{0.5cm} &  $\hat a_V$  & $\hspace{0.5cm} |\hat a_A|$ \hspace{0.5cm}  & $\hat a_V$ & \hspace{0.5cm} $|\hat a_A|$ \hspace{0.5cm} \\
\hline
50  &  $0\%$   & [-0.1863, 0.1821]    &    0.1845  &  [-0.1612, 0.1570] &  0.1591  \\
50  &  $3\%$   & [-0.1872, 0.1830]    &    0.1854  &  [-0.1625, 0.1584] &  0.1605  \\
50  &  $5\%$   & [-0.1887, 0.1845]    &    0.1869  &  [-0.1648, 0.1377] &  0.1628  \\
\hline
100 &  $0\%$   & [-0.1570, 0.1528]    &    0.1551  &  [-0.1358, 0.1317] &  0.1338  \\
100 &  $3\%$   & [-0.1585, 0.1542]    &    0.1566  &  [-0.1381, 0.1339] &  0.1360  \\
100 &  $5\%$   & [-0.1609, 0.1567]    &    0.1591  &  [-0.1418, 0.1567] &  0.1397  \\
\hline
300 &  $0\%$   & [-0.1198, 0.1156]    &    0.1179  &  [-0.1037, 0.0996] &  0.1017  \\
300 &  $3\%$   & [-0.1230, 0.1188]    &    0.1211  &  [-0.1085, 0.1044] &  0.1065  \\
300 &  $5\%$   & [-0.1282, 0.1239]    &    0.1263  &  [-0.1158, 0.1117] &  0.1138  \\
\hline
500 &  $0\%$   & [-0.1057, 0.1015]    &    0.1037  &  [-0.0915, 0.0874] &  0.0895  \\
500 &  $3\%$   & [-0.1103, 0.1061]    &    0.1083  &  [-0.0983, 0.0942] &  0.0963  \\
500 &  $5\%$   & [-0.1172, 0.1130]    &    0.1153  &  [-0.1077, 0.1036] &  0.1057  \\
\hline
1000 & $0\%$   & [-0.0892, 0.0850]    &    0.0872  &  [-0.0773, 0.0732] &  0.0752  \\
1000 & $3\%$   & [-0.0965, 0.0923]    &    0.0945  &  [-0.0877, 0.0836] &  0.0857  \\
1000 & $5\%$   & [-0.1063, 0.1021]    &    0.1044  &  [-0.1002, 0.0961] &  0.0981  \\
\hline\hline
\end{tabular}
\end{center}
\end{table}

% Tabla 6
\begin{table}[!ht]
\caption{Sensitivity on the $\hat a_V$ magnetic moment and the $\hat a_A$ electric dipole moment of the top-quark
through the process $e^-p \to e^-\gamma p \to \bar t \nu_e b p$.}
\begin{center}
 \begin{tabular}{|cccccc|}
\hline\hline
\multicolumn{6}{|c|}{ $\sqrt{s}=10$ TeV,  \hspace{1cm}  $90\%$ C.L.}\\
 \hline
 & & \multicolumn{2}{|c|}{$P_{e^-}=0\%$} & \multicolumn{2}{|c|}{$P_{e^-} = -80\%$}\\
 \hline
 \cline{1-6} \hspace{0.5cm} ${\cal L}\, (fb^{-1})$  & \hspace{0.5cm} $\delta_{sys}$ \hspace{0.5cm} &  $\hat a_V$  & $\hspace{0.5cm} |\hat a_A|$ \hspace{0.5cm}  & $\hat a_V$ & \hspace{0.5cm} $|\hat a_A|$ \hspace{0.5cm} \\
\hline
50  &  $0\%$   & [-0.2108, 0.2066]    &    0.2090  & [-0.1823, 0.1782] &  0.1803 \\
50  &  $3\%$   & [-0.2118, 0.2075]    &    0.2100  & [-0.1838, 0.1797] &  0.1818 \\
50  &  $5\%$   & [-0.2135, 0.2093]    &    0.2117  & [-0.1864, 0.1823] &  0.1844 \\
\hline
100 &  $0\%$   & [-0.1776, 0.1734]    &    0.1757  & [-0.1536, 0.1495] &  0.1516 \\
100 &  $3\%$   & [-0.1792, 0.1750]    &    0.1774  & [-0.1561, 0.1520] &  0.1541 \\
100 &  $5\%$   & [-0.1820, 0.1778]    &    0.1802  & [-0.1603, 0.1562] &  0.1583 \\
\hline
300 &  $0\%$   & [-0.1354, 0.1312]    &    0.1335  & [-0.1172, 0.1131] &  0.1152 \\
300 &  $3\%$   & [-0.1391, 0.1348]    &    0.1372  & [-0.1227, 0.1186] &  0.1206 \\
300 &  $5\%$   & [-0.1449, 0.1407]    &    0.1430  & [-0.1309, 0.1268] &  0.1289 \\
\hline
500 &  $0\%$   & [-0.1195, 0.1152]    &    0.1175  & [-0.1034, 0.0993] &  0.1013 \\
500 &  $3\%$   & [-0.1246, 0.1204]    &    0.1227  & [-0.1111, 0.1070] &  0.1091 \\
500 &  $5\%$   & [-0.1325, 0.1283]    &    0.1306  & [-0.1218, 0.1176] &  0.1197 \\
\hline
1000 & $0\%$   & [-0.1008, 0.0966]    &    0.0988  & [-0.0873, 0.0832] &  0.0852 \\
1000 & $3\%$   & [-0.1090, 0.1048]    &    0.1071  & [-0.0991, 0.0950] &  0.0970 \\
1000 & $5\%$   & [-0.1202, 0.1159]    &    0.1183  & [-0.1132, 0.1091] &  0.1112\\
\hline\hline
\end{tabular}
\end{center}
\end{table}

% Tabla 7
\begin{table}[!ht]
\caption{Sensitivity on the $\hat a_V$ magnetic moment and the $\hat a_A$ electric dipole moment of the top-quark
through the process $e^-p \to e^-\gamma p \to \bar t \nu_e b p$.}
\begin{center}
 \begin{tabular}{|cccccc|}
\hline\hline
\multicolumn{6}{|c|}{ $\sqrt{s}=10$ TeV,  \hspace{1cm}  $95\%$ C.L.}\\
 \hline
 & & \multicolumn{2}{|c|}{$P_{e^-}=0\%$} & \multicolumn{2}{|c|}{$P_{e^-} = -80\%$}\\
 \hline
 \cline{1-6} \hspace{0.5cm} ${\cal L}\, (fb^{-1})$  & \hspace{0.5cm} $\delta_{sys}$ \hspace{0.5cm} &  $\hat a_V$  & $\hspace{0.5cm} |\hat a_A|$ \hspace{0.5cm}  & $\hat a_V$ & \hspace{0.5cm} $|\hat a_A|$ \hspace{0.5cm} \\
\hline
50  &  $0\%$   & [-0.2600, 0.2558]    &    0.2583  & [-0.2248, 0.2207] &  0.2228  \\
50  &  $3\%$   & [-0.2612, 0.2570]    &    0.2595  & [-0.2267, 0.2225] &  0.2246  \\
50  &  $5\%$   & [-0.2633, 0.2591]    &    0.2616  & [-0.2299, 0.2258] &  0.2279  \\
\hline
100 &  $0\%$   & [-0.2190, 0.2147]    &    0.2172  & [-0.1893, 0.1852] &  0.1873  \\
100 &  $3\%$   & [-0.2210, 0.2168]    &    0.2192  & [-0.1925, 0.1883] &  0.1904  \\
100 &  $5\%$   & [-0.2245, 0.2202]    &    0.2227  & [-0.1976, 0.1935] &  0.1956  \\
\hline
300 &  $0\%$   & [-0.1669, 0.1627]    &    0.1650  & [-0.1444, 0.1402] &  0.1423  \\
300 &  $3\%$   & [-0.1714, 0.1671]    &    0.1695  & [-0.1511, 0.1470] &  0.1491  \\
300 &  $5\%$   & [-0.1786, 0.1743]    &    0.1768  & [-0.1613, 0.1572] &  0.1593  \\
\hline
500 &  $0\%$   & [-0.1471, 0.1429]    &    0.1452  & [-0.1273, 0.1232] &  0.1253  \\
500 &  $3\%$   & [-0.1535, 0.1493]    &    0.1517  & [-0.1368, 0.1327] &  0.1348  \\
500 &  $5\%$   & [-0.1633, 0.1590]    &    0.1614  & [-0.1500, 0.1459] &  0.1480  \\
\hline
1000 & $0\%$   & [-0.1241, 0.1198]    &    0.1221  & [-0.1074, 0.1033] &  0.1053  \\
1000 & $3\%$   & [-0.1342, 0.1300]    &    0.1323  & [-0.1220, 0.1179] &  0.1199  \\
1000 & $5\%$   & [-0.1480, 0.1438]    &    0.1462  & [-0.1394, 0.1353] &  0.1374  \\
\hline\hline
\end{tabular}
\end{center}
\end{table}

%\newpage

\vspace{2cm}

\begin{center}
{\bf Acknowledgments}
\end{center}

A. G. R. and M. A. H. R. acknowledge support from SNI and PROFOCIE (M\'exico).

%\pagebreak

\pagebreak

\begin{figure}[t]
\centerline{\scalebox{0.8}{\includegraphics{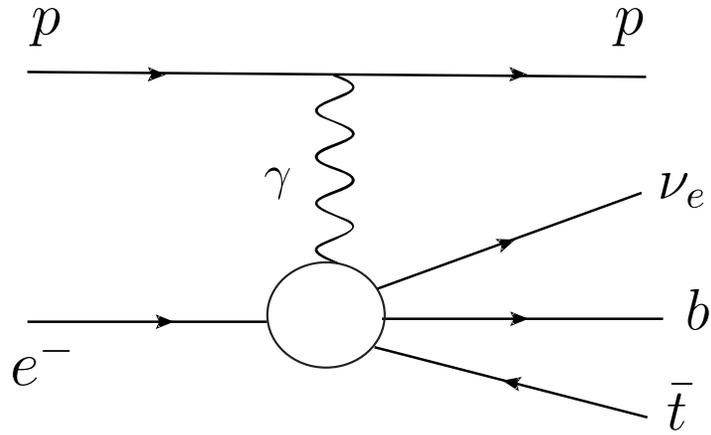}}}
\caption{ \label{fig:gamma1} A schematic diagram for the process
$e^-p \to e^-\gamma p \to \bar t \nu_e b p$.}
\label{Fig.1}
\end{figure}

\begin{figure}[t]
\centerline{\scalebox{0.8}{\includegraphics{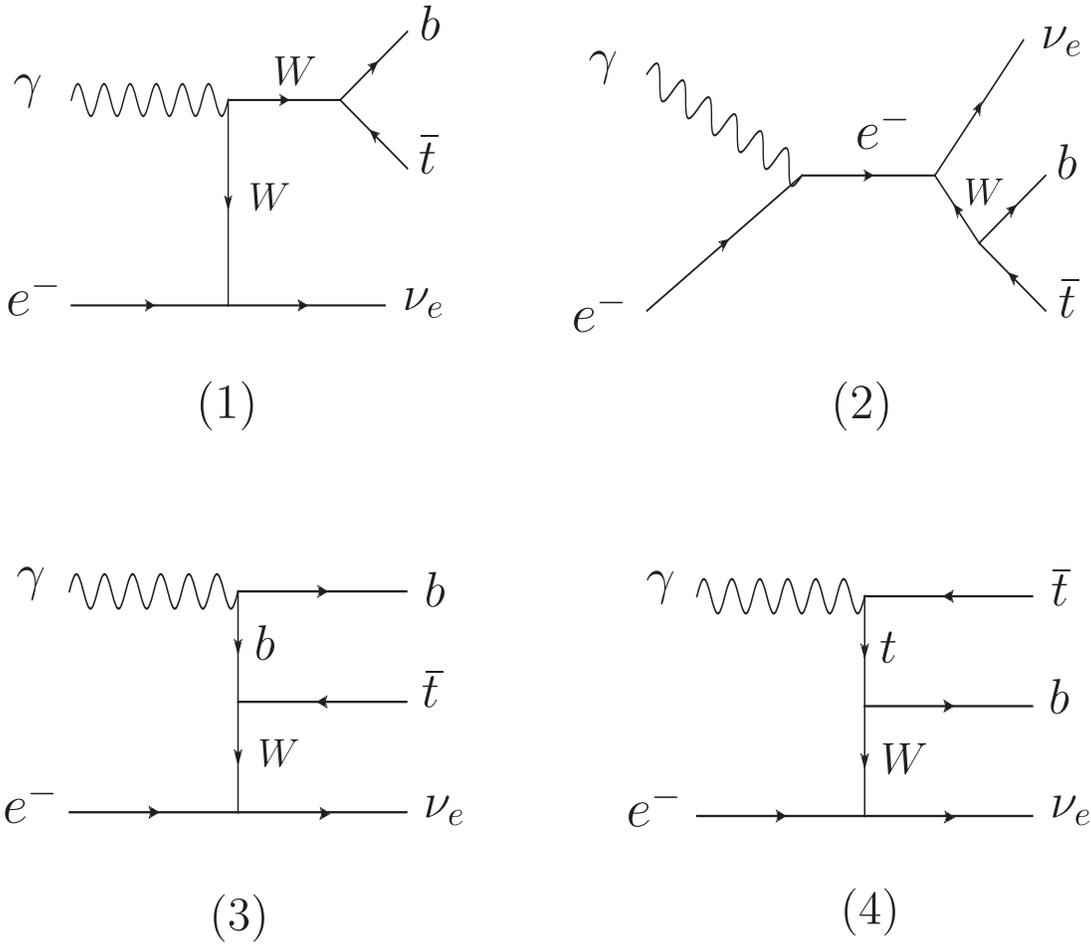}}}
\caption{ \label{fig:gamma2} Feynman diagrams contributing to the subprocess
$e^-\gamma \to \bar t \nu_e b $.}
\label{Fig.2}
\end{figure}

\begin{figure}[t]
\centerline{\scalebox{1.2}{\includegraphics{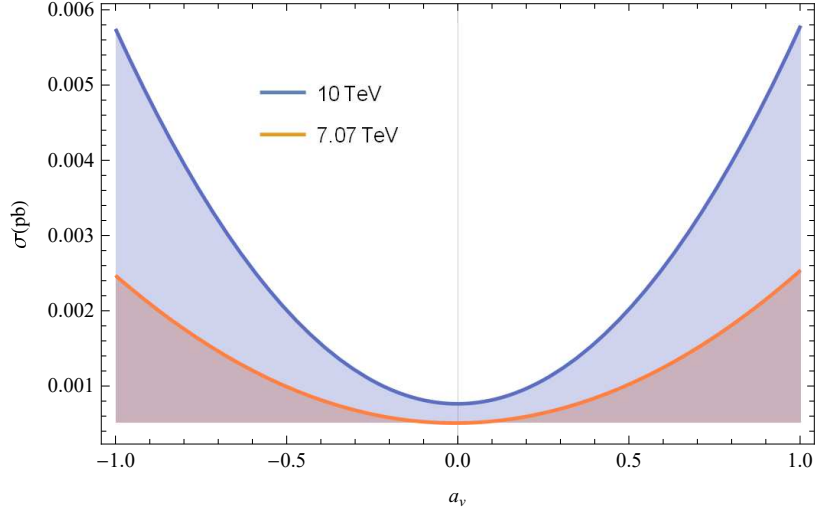}}}
\caption{ The total cross sections of the process
$e^-p \to e^-\gamma p \to \bar t \nu_e b p$ as a function of $\hat a_V$
for center-of-mass energies of $\sqrt{s}=7.07, 10$\hspace{0.8mm}$TeV$ at the FCC-he.}
\label{Fig.3}
\end{figure}

\begin{figure}[t]
\centerline{\scalebox{1.2}{\includegraphics{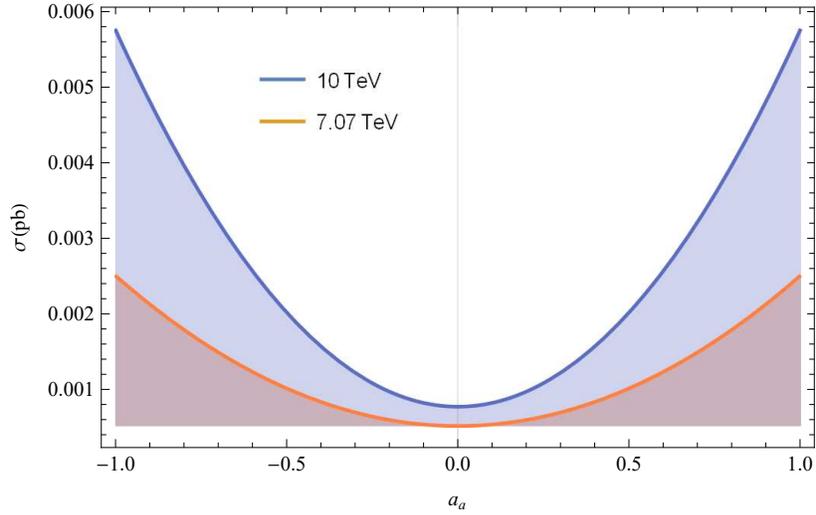}}}
\caption{ Same as in Fig. 3, but for $\hat a_A$.}
\label{Fig.4}
\end{figure}

\begin{figure}[t]
\centerline{\scalebox{1.2}{\includegraphics{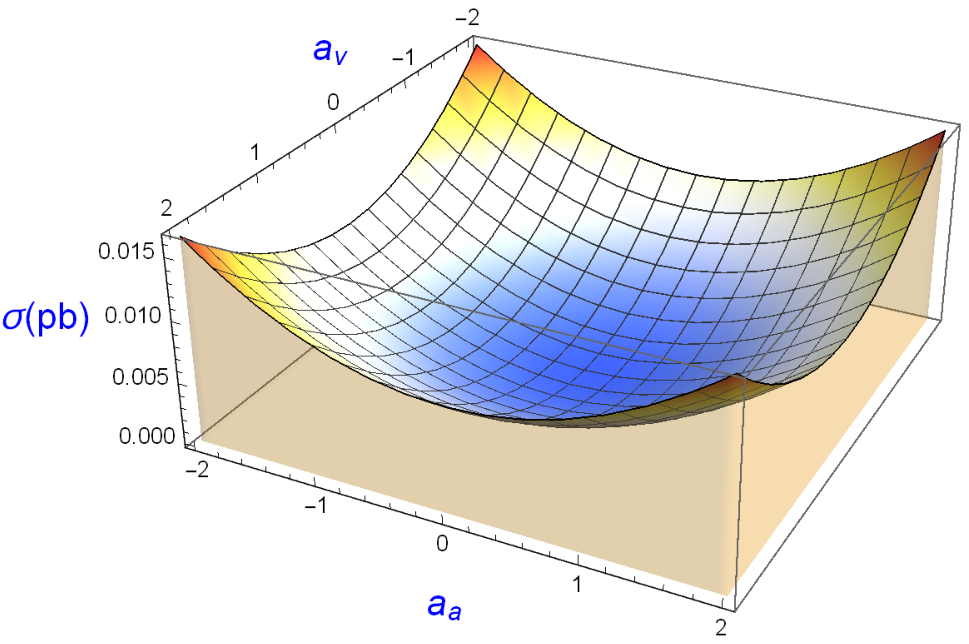}}}
\caption{ \label{fig:gamma1} The total cross sections of the process
$e^-p \to e^-\gamma p \to \bar t \nu_e b p$ as a function of $\hat a_V$ and $\hat a_A$
for center-of-mass energy of $\sqrt{s}=7.07$\hspace{0.8mm}$TeV$ at the FCC-he.}
\label{Fig.5}
\end{figure}

\begin{figure}[t]
\centerline{\scalebox{1.2}{\includegraphics{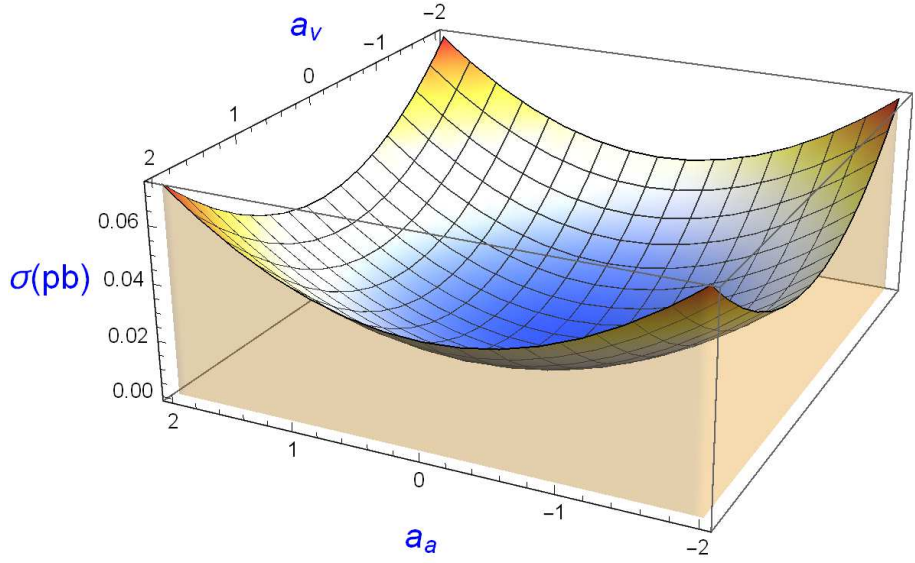}}}
\caption{ \label{fig:gamma2} Same as in Fig. 5, but for center-of-mass energy of
$\sqrt{s}=10$\hspace{0.8mm}$TeV$.}
\label{Fig.6}
\end{figure}

\begin{figure}[t]
\centerline{\scalebox{1.1}{\includegraphics{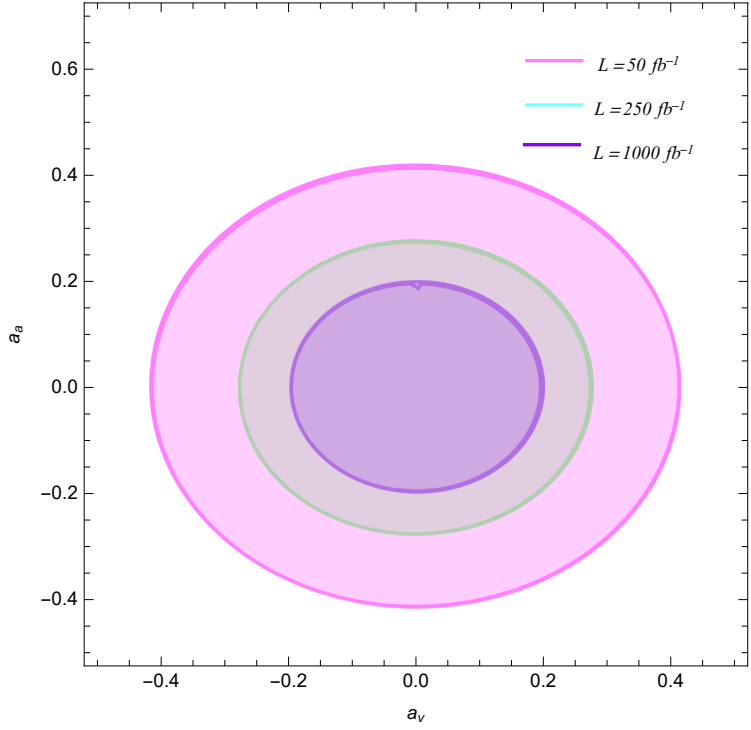}}}
\caption{ \label{fig:gamma1x} Sensitivity contours at the $95\% \hspace{1mm}C.L.$ in the
$\hat a_V-\hat a_A$ plane through the process $e^-p \to e^-\gamma p \to \bar t \nu_e b p$
for $\sqrt{s}=7.07$\hspace{0.8mm}$TeV$ at the FCC-he.}
\label{Fig.7}
\end{figure}

\begin{figure}[t]
\centerline{\scalebox{1.1}{\includegraphics{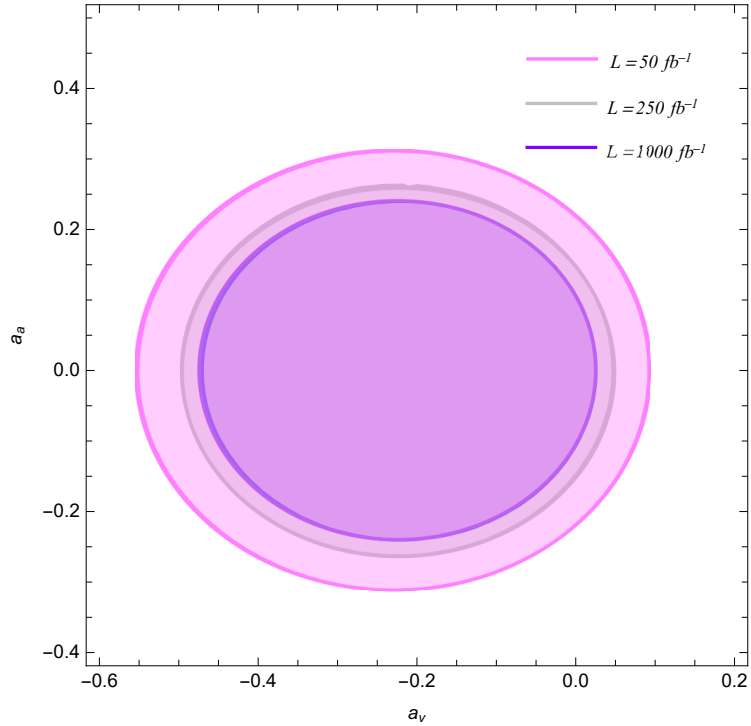}}}
\caption{ \label{fig:gamma2x} Same as in Fig. 7, but for $\sqrt{s}=10$\hspace{0.8mm}$TeV$.}
\label{Fig.8}
\end{figure}

\begin{figure}[t]
\centerline{\scalebox{1.2}{\includegraphics{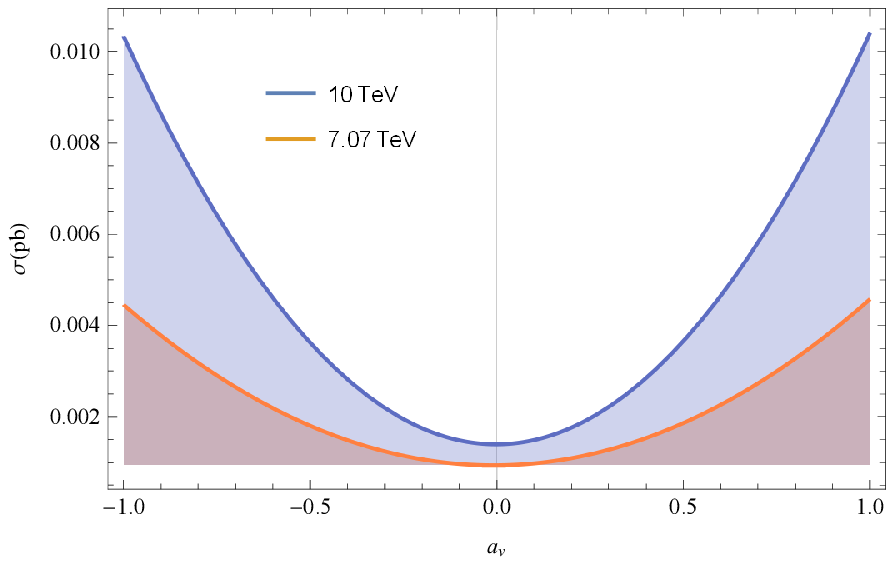}}}
\caption{ \label{fig:gamma15} Same as in Fig. 3, but with polarized beams $P_{e^-}=-80\%$.}
\label{Fig.6}
\end{figure}

\begin{figure}[t]
\centerline{\scalebox{1.2}{\includegraphics{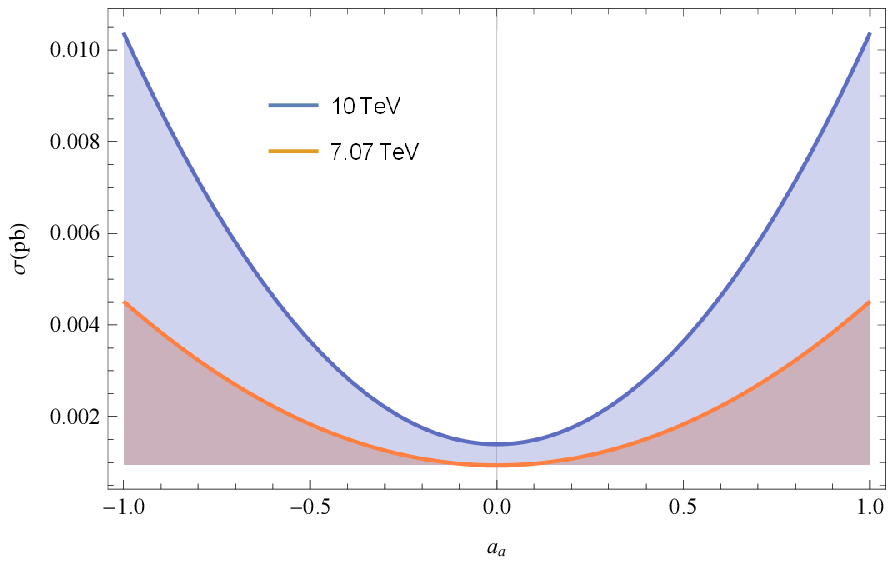}}}
\caption{ \label{fig:gamma6} Same as in Fig. 4, but with polarized beams $P_{e^-}=-80\%$.}
\label{Fig.7}
\end{figure}

\begin{figure}[t]
\centerline{\scalebox{1.2}{\includegraphics{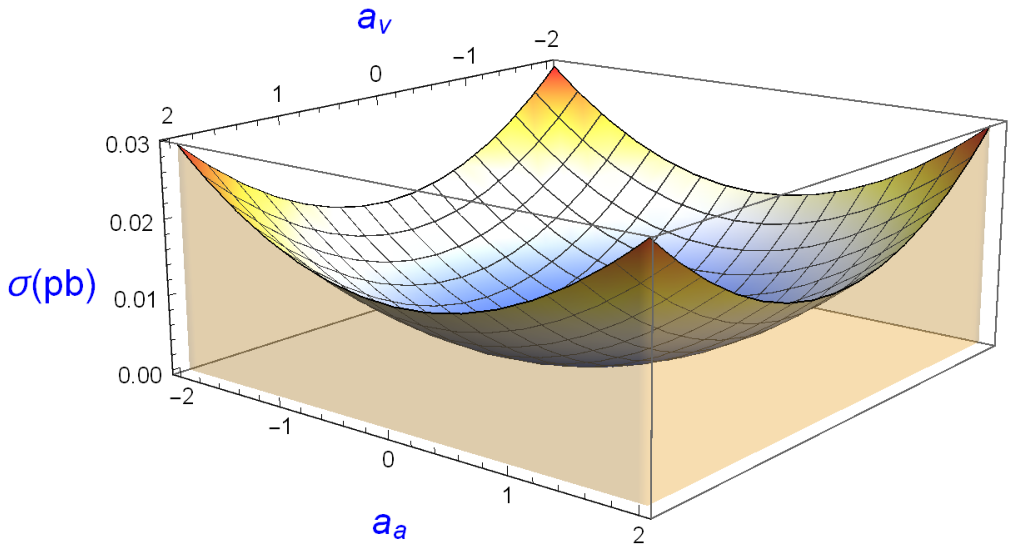}}}
\caption{ \label{fig:gamma6x} Same as in Fig. 5, but with polarized beams $P_{e^-}=-80\%$.}
\label{Fig.8}
\end{figure}

\begin{figure}[t]
\centerline{\scalebox{1.2}{\includegraphics{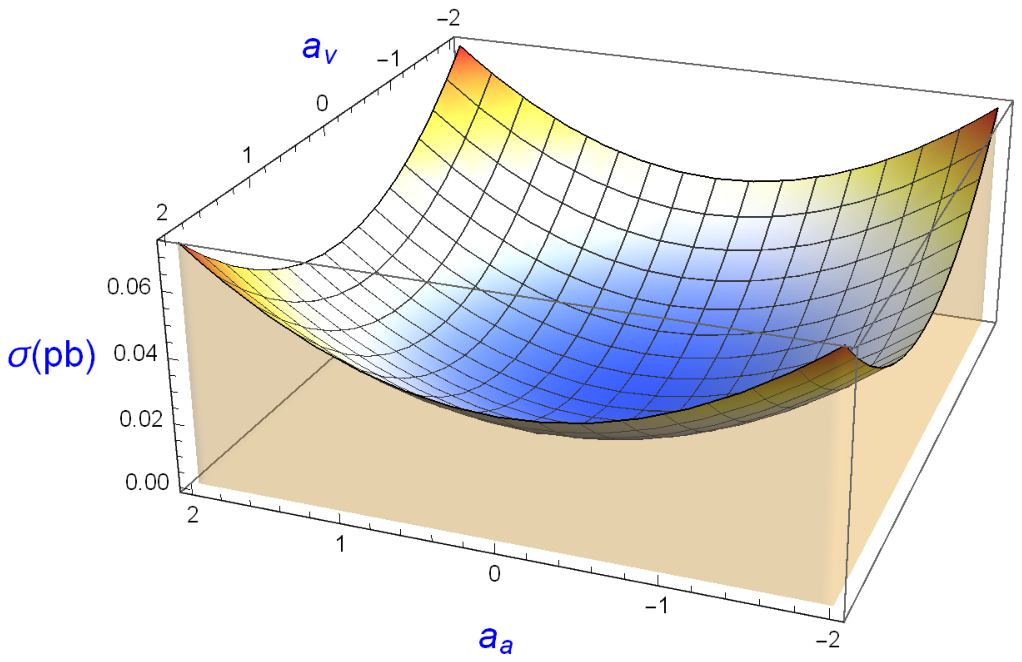}}}
\caption{ \label{fig:gamma6x} Same as in Fig. 6, but with polarized beams $P_{e^-}=-80\%$.}
\label{Fig.8}
\end{figure}

\begin{figure}[t]
\centerline{\scalebox{1.1}{\includegraphics{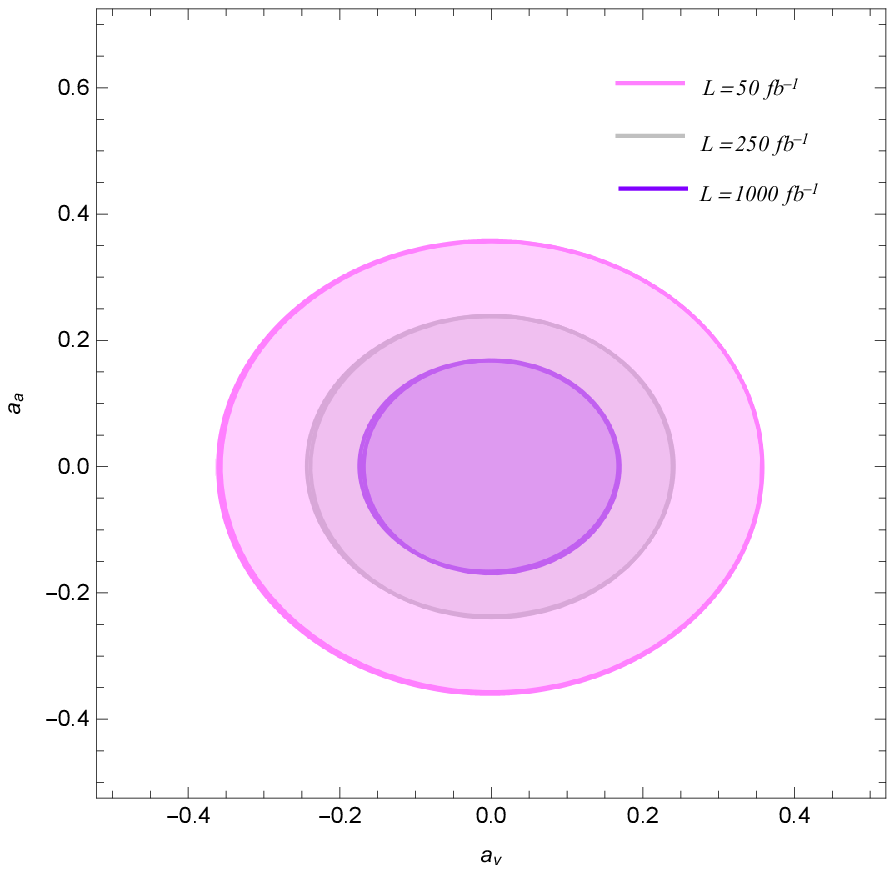}}}
\caption{ \label{fig:gamma6x} Same as in Fig. 7, but with polarized beams $P_{e^-}=-80\%$.}
\label{Fig.8}
\end{figure}

\begin{figure}[t]
\centerline{\scalebox{1.1}{\includegraphics{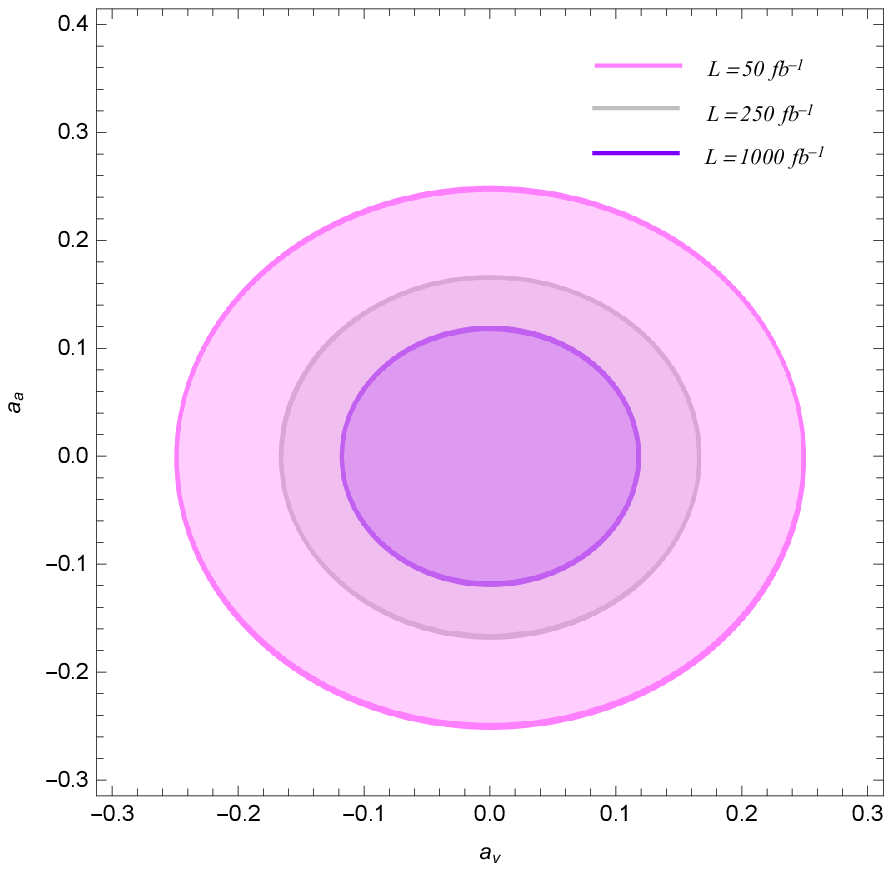}}}
\caption{ \label{fig:gamma6x} Same as in Fig. 8, but with polarized beams $P_{e^-}=-80\%$.}
\label{Fig.8}
\end{figure}

\end{document}